\documentclass{article} 

\usepackage[utf8]{inputenc}

\usepackage{iclr2018_workshop,times}
\usepackage{url}
\usepackage{graphicx}
\usepackage{multirow}
\usepackage{lipsum}
\usepackage{amsthm}
\usepackage{graphicx}

\usepackage{hyperref}
\hypersetup{
    colorlinks=true,
    urlcolor=[rgb]{0.188, 0.498, 0.886},
    linkcolor=[rgb]{0.188, 0.498, 0.886},    
    filecolor=magenta,      
    citecolor=[rgb]{0.188, 0.498, 0.886},
}

\usepackage{algorithm} 
\usepackage{program} 
\usepackage{algorithmic}
\usepackage{listings}
\usepackage{geometry}

\usepackage{caption,setspace}
\DeclareCaptionFormat{listing}{\rule{\dimexpr0.9\columnwidth+17pt\relax}{0.4pt}\par\vskip1pt#1#2#3}


\DeclareMathVersion{sans}
\SetSymbolFont{operators}{sans}{OT1}{cmbr}{m}{n}
\SetSymbolFont{letters}  {sans}{OML}{cmbrm}{m}{it}
\SetSymbolFont{symbols}  {sans}{OMS}{cmbrs}{m}{n}

\lstnewenvironment{sflisting}[1][]
  {\lstset{#1}\mathversion{sans}}{}

\usepackage[normalem]{ulem}

\definecolor{grayalias}{HTML}{3F4444}

\definecolor{bluealias}{HTML}{307FE2}

\title{Robot hazards: from safety to security}

\author{
  Laura Alzola Kirschgens,
  \textbf{Irati Zamalloa Ugarte},
  \textbf{Endika Gil Uriarte},\\
  \textbf{Aday Muñiz Rosas},
  \textbf{Víctor Mayoral Vilches}\\
  Alias Robotics S.L. \\
  Vitoria-Gasteiz, Álava, Spain \\
  \texttt{laura@aliasrobotics.com} \\
}

\usepackage{pdfpages}

\begin{document}

\maketitle
\vspace{-1em}
\begin{abstract}
Robotics landscape is experiencing big changes. Robots are spreading and will soon be everywhere. Systems traditionally employed in industry are being replaced by collaborative robots, while more and more professional and consumer robots are introduced in people's daily activities. Robots are increasingly intertwined with other facets of IT and envisioned to get much more autonomy, interacting physically with humans. We claim that, following Personal Computers (PCs) and smart-phones, robots are the next technological revolution and yet, robot security is being ignored by manufacturers. The present paper aims to alert about the need of dealing not only with safety but with robot security from the very beginning of the forthcoming technological era. We provide herein a document that reviews robot hazards and analyzes the consequences of not facing these issues. We advocate strongly for a security-first approach as a must to be implemented now.

\end{abstract}




\section{Introduction}
\label{sec:intro}

Robots are not new: they have played a big role in manufacturing, medicine, warehouse operations, and other industries for years. However, robotics, as a field, is experiencing big changes. Robots are morphing due to a confluence of mighty processing power, artificial intelligence, natural language processing and exponential data growth. These machines are spreading and will soon be everywhere. Robots traditionally employed in industry are being replaced by collaborative robots, while more and more professional and consumer robots are being introduced in our daily activities. They are already present at medical and health institutions, airports, offices, shopping centers, restaurants and many homes. Moreover, robotics is becoming increasingly intertwined with facets of IT such as the cloud, mobile devices and the Internet of Things (IoT). And, unlike traditional robots, the coming generation of these machines is being envisioned and designed to gain more autonomy. There is a growing consensus on that advances in AI will enable robots to move freely in unstructured environments.

It was only back in the 90's when both personal computers and the use of internet was democratized and those technologies' insecurities became more than evident for the general public: Neither the internet nor computers where conceived to be secure. Suddenly, producers had to admit their products' vulnerabilities and flaws and find mechanisms to protect them from an increasing array of external threats. Right until then, companies had rushed their products to the market taking advantage of the hype but without paying any attention to the forecoming consequences. Something very similar is happening now with robots. The growth of certain robots such as collaborative robots (cobots) in industry\footnote{And soon to expand to other market sectors.} resembles other past disruptive changes. Based on our observations, we claim that following PCs and smartphones, cobots are the next technological revolution. Yet, besides its similarities to the PC revolution, robots present much higher risks for their users. Plainly explained: when a PC is hacked, the output damage usually remains virtual and, although linked to reality in many aspects, the direct consequence of the breach generally stays non-material. Meanwhile, when a robot vulnerability is exploited, apart from this privacy violations, data or economic losses, there is another major effect to be considered: physical outcomes by robot malfunction. Robots can harm people and things. And this is why there is an urgent need for rethinking how we protect robots. 

Robots have to be protected as they go mainstream, connected and free. In order to be secure, reliable and safe. Nowadays, experts across both the robotics and cybersecurity fields do admit that security issues are already on the table of internal discussion among manufacturers and end-users of robots. However, robot makers take the chance of a fast-growing market and rush their products into it without giving an adequate consideration to security. Not enough attention is yet being given to well-known security issues that did already prove to be devastating at the edge of other technological revolutions, such as the spread of commercial computer networks. 

Without rushing into any kind of Skynet, we find it extremely necessary to create public awareness on the need of dealing not only with safety but increasingly with security of robots, from the very beginning of this robotic revolution. In order to do so, we provide herein a complete review of the most relevant robotics malfunctions, analyze downstream societal implications of those and provide a conclusion. The following Section \ref{sec:doc} documents a series of different cases of robots causing trouble in the industrial, professional and consumer sectors over the last decades, including current issues. Section \ref{sec:consecuences}, analyses the human, economical, legal and corporate image consequences that safety and security problems cause. Finally, in Section \ref{sec:conclusion}, we conclude this work discussing our viewpoint for future developments in robotics regarding security.

\section{Documentation} 
\label{sec:doc}

Robots have worked in industry for a long time. The first installation of a cyber-physical system in a manufacturing plant was back in 1962 \cite{historyofrobotics}. Mostly employed in manufacturing, and particularly present in the car industry, for decades, these machines were originally designed to carry out repetitive, dangerous or dull tasks. To be accepted as part of the working areas and procedures where humans were active too, it soon became a must for them to fulfill safety standards. Companies cared about the way these machines affected their environment and surroundings; and safety concerns were mainly approached by getting them to act physically separated from humans, in order to avoid the risky contact among both. Now, along with the new free-roaming, connected and omnipresent robots, there are new concerns rising. Influential opinions, the general public and companies increasingly worry about what might happen after a robot suddenly went rogue, for example, due to a vulnerability in their system being exploited by hackers. In this section we document a series of different cases of robots causing trouble in the traditional industrial sector over the last decades, as well as in the newer industrial, professional and consumer sectors, including current issues.

\begin{figure}[h!]
\centering
 \includegraphics[width=\textwidth]{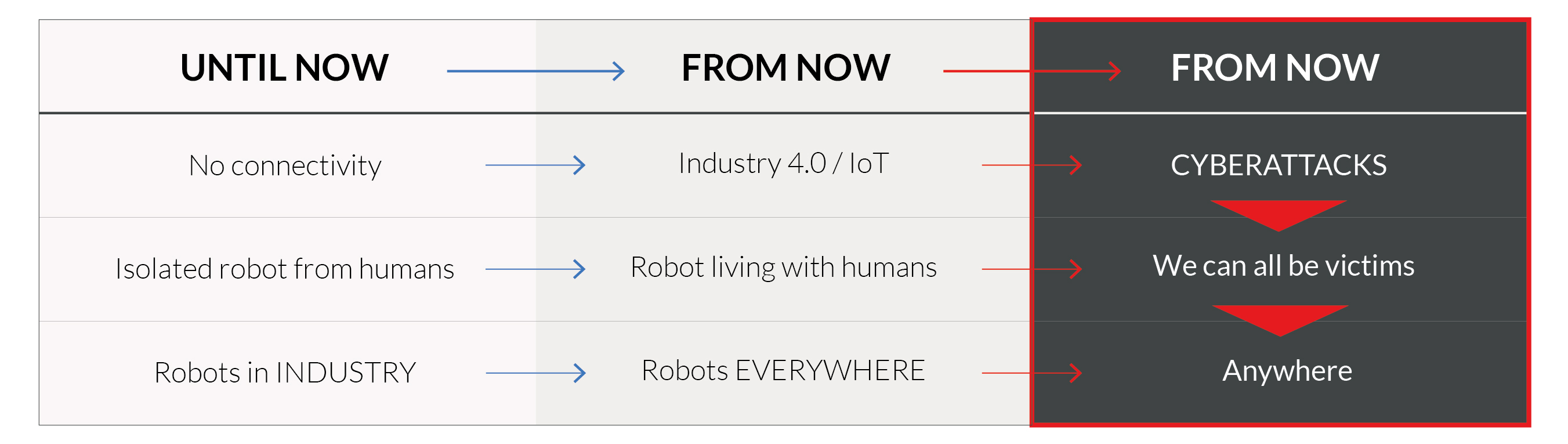}
\caption{\footnotesize Today we are facing an upward trend of connectivity in most moments of our lives including robot-human and robot-robot interactions, what makes them a hot spot for cyber attacks. While robots leave industry to live and interact with humans, all of us start to be increasingly exposed to greater risks caused by hacked robots anywhere.}
\label{fig:comparation}
\end{figure}
\subsection{From the eighties until nowadays: traditional robots in industry}
 
The first time an industrial robot killed a human was in 1979. Robert Williams, a 25-year-old factory worker in a Ford Motor Company casting plant in Flat Rock, Michigan, was asked to scale a massive shelving unit to manually count the parts in there \cite{firstkiller}. While Williams was up, a robot arm tasked with parts retrieval silently came upon the young man, striking him in the head and killing him instantly. The machine kept on working, but Williams laid death on the floor for 30 minutes before his body was found by concerned co-workers, according to the Knight-Ridder newspaper's reporting. There were not any fences in place to protect Williams. No alarms notified him of the approaching arm, and no technology could alter the robot's behavior in the presence of a human. In 1981 and under similar circumstances, the next death by a robot occurred, this time in Japan \cite{deathjapan}. Again, a robot arm failed to sense a man and accidentally pushed him to his death. 37-year-old Kenji Urada was trapped by the robot which pinned him against a machine with cut gears. The accident happened at a plant of Kawasaki heavy industries, and, according to factory officials, this time there were safety measures in place. A wire mesh fence around the robot would have shut off the unit’s power supply when unhooked. But instead of opening it, Urada had apparently jumped over the fence and the claw of the robot pushed him against the machine tooling device.

These two cases were the beginning of a horror story starred by killer robots which lasts until these days. Decades later after the beginning of the use of robots in industry, reports of human deaths caused by robots feel even more commonplace. Although many fences and safety measures are in place, dramatic accidents still prevail. According to data from 2014 distributed by the Occupational Safety and Health Administration from the United States of America \cite{33death}, industrial robots caused at least 33 workplace deaths and injuries in that country during the preceding 30 years, this is, since the beginning of the eighties. In 2015, a robot killed a man in a Volkswagen plant in Germany \cite{threat2}. Two months later, in August that same year, another man was killed by a robot at a car parts factory in India \cite{indiakill}. And, one year later, in 2016, in Alabama, an auto parts supplier had to pay a 2.5 million dollar fine after a 20 year old woman, Regina Allen Elsea, at the age of 20, was crushed to death by a robot at the Ajin USA plant \cite{Alabama}. After the investigation of this last death, the Occupational Safety and Health Administration announced 23 safety violations issue. 

Studies underline that the described cases are not exceptional. In fact, there are several recent reports stating that industrial robots are linked to thousands of accidents per year around the world. A work published in 2013, for example, stated that, only in Germany, more than 100 accidents related to industrial are registered every year \cite{hundredgermany}. Moreover, while, at least officially, most of the incidents happening in traditional industrial robots are categorized as "accidents", they demonstrate the potential consequences of malfunctions in these machines.

\subsection{From now on: free-roaming and connected robots spreading everywhere}
There is a growing consensus that, in the next decade, advances in AI will make it increasingly possible for robots to freely move in unstructured environments. On the other hand, robotics is becoming intertwined with facets of IT such as the cloud, mobile devices and the Internet of Things (IoT). Related to these contact, further than only thinking of how the machine may affect the environment (safety), there is a new concern on how external factors can affect the machine itself first, this is, change its behavior, leading to a situation were the robot may (re)act in a negative way. As Giaretta  \emph{et al.} outline \cite{2018arXiv180504101G}, "a hacked robot, used for instance in a private home or even worse in a public space, like an airport, can have tremendous consequences for the safety of human beings, especially when it is a breeze to remotely turn it into a 'cyber and physical weapon', exposing malicious behaviours". 
\subsubsection{"New" industrial robots, collaborative robots}

In the last years, as outlined in the introduction, a new kind of robot is spreading in manufacturing plants.
Unlike traditional robots, the coming generation of these machines is being envisioned and designed to get much more autonomy and operate alongside humans. These new industrial robots that are being featured at the core of Industry 4.0 are the collaborative robots, commonly known as cobots \cite{wordstomorrow}. Cobots are robots intended to interact physically with humans in a shared work-space and are typically aimed at complementing and augmenting human capabilities to perform tasks in a repetitive and more efficient manner. 

The growth of these robots on industrial applications resembles other past disruptive changes and provides a direct answer to what many are already demanding: a switch from mass manufacturing to mass customization. Most of these robots provide easy-to-configure mechanisms via simplified user interfaces that abstract the final user from complex programming languages. Yet, although these simple mechanisms reduce the probability of operating these machines unsafely, they are not error-free. As Cerrudo and Apa describe at \cite{hackingbeforeskynet, hackingbeforeskynet2}, when analyzing robots from Universal Robots, one of the most popular cobots brands, robots are easily re-purposed in a malicious manner with a simple buffer overflow applied to the Modbus external port that the robot provides. Furthermore, as publicly advertised by Cerrudo and Apa, no apparent action has been taken by the manufacturer to prevent this from continuously happening. And this is not an uncommon behavior. Several robot manufacturers employ networks of distributors, re-sellers and integrators to help them spread their businesses and brands faster. This strategy is currently working specially well for cobots, priced an order of magnitude below their traditional industrial counterparts. However, with these adjusted margins\footnote{That apparently doesn't even allow for security patches.} and aggressive business strategies based on growth and deployment, it seems reasonable to ask: thousands of insecure cobots are being deployed all around the world to \emph{collaborate} with humans, who will be held responsible when these security holes get exploited and cause damages? How would Universal Robots or other manufacturers respond when these reported, non-patched and public vulnerabilities cause hazards? Would they train hundreds of distributors all around the world on how to patch basic security wholes manually? Or maybe, would they consider security practices in the development and deployment cycle of these robots?

Collaborative robots are here to stay. Based on our observations, we claim that following after Personal Computers (PCs) and smartphones, cobots are the next technological revolution that will change many industrial operations. And yet, how long until the insecurity of these devices causes damage that forces manufacturers to act? Would robot companies acknowledge that security has critical implications for safety? These new robots in industry will be connected to network, being more vulnerable to be hacked. And, in an industrial setting, a hack meant to simply disrupt a system could end up affecting the quality of an entire line of product or even provoke the halting of a manufacturing run completely, costing millions of dollars. Moreover, a hacked robot in a business or industrial setting could also be used to access other robots sharing the same network and configuration. As stated by Cerrudo and Apa from IOActive \cite{hackingbeforeskynet}, in those circumstances, a hacked robot can become an "attack platform to exploit vulnerabilities in other network devices and propagate the attack". 

Like Universal Robots, many robotic companies have indeed a complicated landscape. These collaborative capabilities, while for now, mainly, developed and applied in industry will soon reach other areas and applications. The principles of collaborative robots will soon be reused in other areas such as health-care as well as in a wide range of professional tasks involving cleaning, cooking or storing. All connected.

\subsubsection{Professional robots}

A professional robot  is a service robot used for a commercial task, usually operated by a properly trained operator \cite{ifr}. This kind of robots are spreading in medical and health institutions, airports, offices, shopping malls, restaurants. Here too, along with these free-roaming, internet connected robots, new concerns are rising. General public and companies worry about what may happen if a robot suddenly spins out of control, namely, due to a vulnerability in their system being exploited by hackers. There is a new concern on how external factors can affect the machine, leading to a situation where the robot may act in a negative way, for example by executing other movements than expected or by leaking data. 

In this segment, \emph{surgical robots} have now taken a relevant position. For instance in medicine, an area where reliance on security seems substantially critical. But this growth is already showing downsides. As a study carried out in 2014 by researchers at the University of Illinois at Urbana-Champaign, the Massachusetts Institute of Technology and Chicago's Rush University Medical Center stated, over a 14 year period (2000-2013), at least 144 deaths and more than 1,000 injuries related to surgical use of robots were detected \cite{threat5}. The events included, among other reasons, uncontrolled movements and spontaneous powering on/off of the machines, electrical sparks unintended charring and damaging accessory covers and burning of body tissues, as well as loss of quality video feeds and/or reports of system error codes.
 
Furthermore, \emph{security surveillance} is another professional niche were collaborative robots are being implanted. And together with some positive aspects, the negatives are also emerging. Back in 2016, for example, a robot acting as a security guard knocked down and injured a toddler for still unknown reasons. And the robot, called Knightscope K5, a five-foot, 300-pound robot that had begun trials in the mall the year before, just kept moving \cite{threat3}. It happened at the Stanford Shopping Center in Silicon Valley. The 16-month-old child was not seriously hurt, but Knightscope obviously would not want something like this to happen again. Yet, the incidents starred by that robot did not stop. One year after a Knightscope K5 “drowned” in a decorative mall-fountain. The machine was unable to avoid falling down the stairs surrounding the water \cite{water}. Furthermore, in 2018, this robot model has been described as "creepy" by different women \cite{creepy} at the Laguardia Airport in New York. Several female passengers said they were irked when the machine rolled up and ogled them, claiming the machine encroached their personal space.

\subsubsection{Consumer sector robot}

Roomba will not be home alone soon. In the next years a greater number of increasingly sophisticated robots are expected to be used for diverse tasks by individual clients, including not only chores, but communication, entertainment or companionship. Consumer robots are personal service robots used for a non-commercial task, usually by lay people \cite{ifr}. Their segment within robotics is growing dazzlingly and therefore getting vendors busy developing their products and pushing them to the market to take advantage. In this rush, security seems to be a lower priority, one that might complicate or increase the cost of their systems. 

Apart from the need of making them secure and privacy respecting, household robots pose other specific challenges to robot manufacturers and researchers.
Back in 2009 already, a team from the University of Washington evaluated \cite{denning2009spotlight} how a hacked service or consumer sector robot could result in a big problem. They tested the security of three consumer-level robots: the WowWee Rovio, a wireless, buglike rolling robot marketed to adults as a home surveillance tool controlled over the internet and including a video camera, microphone and speaker; the Erector Spykee, a toy wireless web-controlled “spy” robot that had a video camera, microphone and speaker; and the WowWee RoboSapien V2, a more dexterous toy robot controlled over short distances using an infrared remote control. The authors identified scenarios in which a robot might physically harm its owner or the home environment.

\subsection{Into a new era of hacked robots}

\begin{figure}[h!]
\centering
 \includegraphics[width=\textwidth]{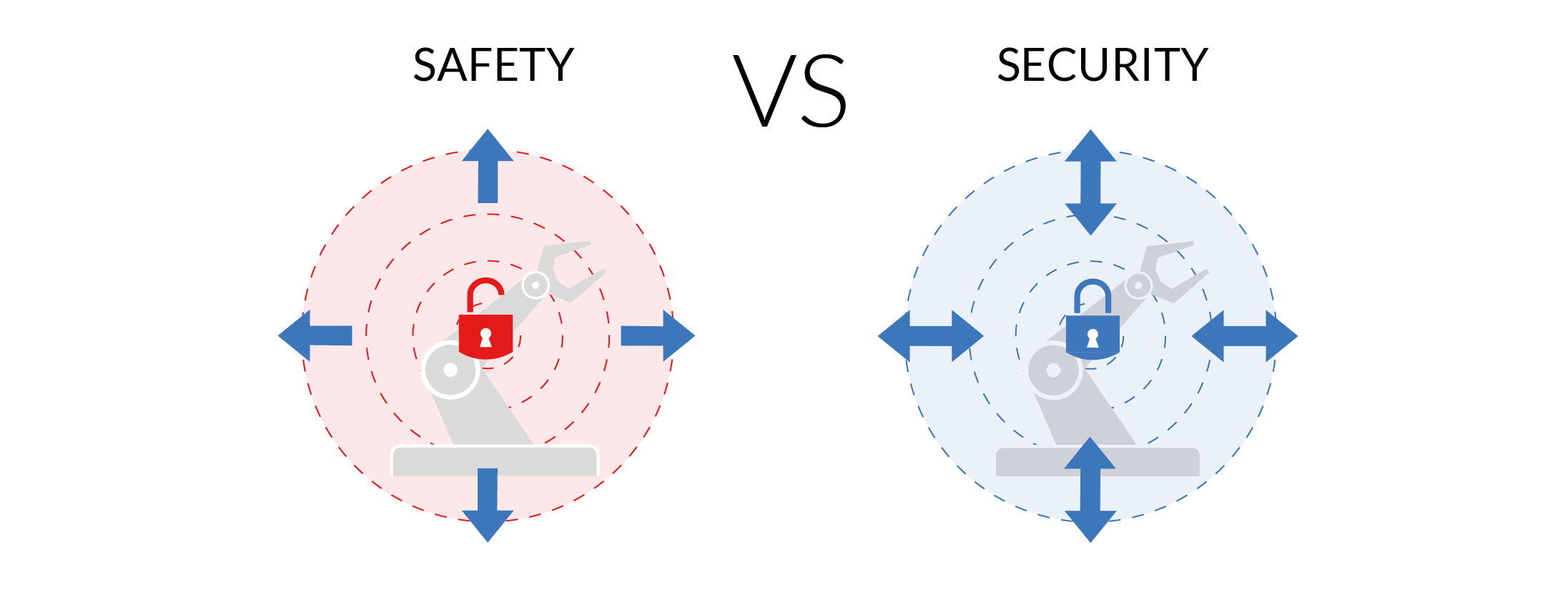}
\caption{\footnotesize Safety cares about the possible damage a robot may cause in its environment, whilst security aims at ensuring that the environment does not disturb the robot operation. Safety and security are connected matters. A security-first approach is a must to ensure safe operations.}
\label{fig:safety-security}
\end{figure}

From industry to consumer robots, going through professional ones, most of these machines are not prepared for cyber-threats and security vulnerabilities. Manufacturers' concerns, as well as existing standards, focus mainly on safety. Security is not being considered as a relevant matter. The architecture of many of these robots is actually the greatest example of how little security has been considered. A simple way to understand this lack of concern appears when looking closely at robot architectures nowadays. Most, include at least two differentiated networks: an external one, meant to be used by end-users for operating the robot, which usually is less protected than it should; and an internal one, where sensors, actuators and other components within the robot exchange information and cooperate putting together what the robot is: a machine composed by distributed hardware and software components governed by a controller that requires access to the information from all these devices. External networks do typically include some sort of encryption, authentication and integrity mechanisms however, internal networks are often unprotected. Not even a simple authentication check is required, what means that anyone with physical access to the robot could potentially access this internal network and disrupt the robot behavior completely.

Safety has been accounted for, partially at least. Unfortunately, the lack of security has safety repercussions. We are about to dive into a new era of hacked robots. Those manufacturers that learn to apply a security-first approach will have a competitive advantage.

\section{Consequences of \emph{not} protecting robots in this new era}
\label{sec:consecuences}
Robots being unprotected, unsafe, insecure and/or not private leads to multiple outcomes, all affecting companies negatively. The cases documented in the previous section serve as a brief glance at the real and practical consequences of not considering security a major issue. Fact is that not enough attention is yet been given to well-known security issues that did already prove to be devastating at the edge of other technological revolutions, such as the spread of commercial computer networks and the internet use. Sadly, nowadays, we see robot makers taking the chance and rushing their products to market without giving adequate consideration to security. Moreover, it is common for manufacturers lacking good security practices to not know how to deal with vulnerability reports. Most of them probably do not even have an effective procedure in place to handle reports, and neither to provide security fixes to customers. If lessons are not learned and robot manufacturers fail to take a security-first approach now, it may haunt them soon. In the following analysis of some of the resulting unfavorable effects, we have chosen three thematic points: Human loss and injuries, data theft and privacy issues, and the destruction of the corporate image. Yet, there is one dimension that intersects with all of them: the economical one.

\subsection{Human loss and injuries}

The effects of not taking safety and security seriously enough are visible and devastating since the very beginning of robots’ implantation in different industrial scenarios back in the eighties. Dead or severely injured humans have been and still are the most dramatic result of unsafe and uncontrolled robots. In terms of business, those are the effects hardest to hide too. Ergo, the most expensive ones, both directly and indirectly. Because, once gone public, the derived huge lawsuits and fines are almost irremediable. Even the first case of a robot killing a man, back in 1979 in Michigan, illustrates this idea, as it was the starting point for the extensive history of fines and lawsuits paid by companies in those cases where incidents with robots reached trials. In William’s case, the jury agreed that not enough care had been put into the design of the robot to prevent a death like this. His family won a 10 million dollar lawsuit for his wrongful death from Unit Handling Systems, the manufacturer that designed the robot \cite{10million}. 

Not investing enough resources, time and effort into protecting workers from robots has costed companies lots of money. The latest of the known and available cases goes back to 2016, \cite{Alabama} when an auto parts supplier had to pay a 2.5 million dollar fine. However, deaths and serious injuries caused by robots are not only constricted to industrial machines. Among the professional sector ones, for example, the \emph{Da Vinci} surgical robot is one of the machines with the most lawsuits. In a 2014 regulatory filing, Intuitive Surgical, its manufacturing company, said they were facing 3,000 product-liability claims over surgeries taking place between 2004 and 2013. Moreover, the firm set aside 67 million dollar to settle an undisclosed number of claims. Furthermore, in the first six months of this year, Intuitive has reported to have put aside 16 million dollar to settle legal claims \cite{davinci}.

\subsection{Data theft and privacy issues}

Following an economical narrative of the consequences of not paying enough attention to security, it needs to be highlighted that companies do also fear loss or theft of information. When a computer stores information, there always is a risk of exposure, accidental or desired, when some hacker or criminal breaks in and steals data. Robots are storing data that could be accessed and stolen. Furthermore, companies could be extorted with this data in exchange. Privacy issues, such as the legal consequences of customers’ data leak or misuse by third parties have also been put on the table recently. Lawsuits filed by clients and end users, or even by manufacturers, will be growing awareness around privacy and its importance rises. Moreover, as it already happened at the beginning of the PC era, robots are introducing complex privacy and security issues that may not have been considered enough yet. While if a PC is hacked, data loss and identity theft are the potential results, robots melding advanced technology with mobile capabilities that could get compromised have the potential of doing serious physical damage to people and property around them. 

\subsection{Destruction of the corporate image}

Corporate image is generally defined as the mental picture that springs up at the mention of a firm's name. It is the public perception of the company, a composite psychological impression that continually changes with the firm's circumstances, media coverage, performance, pronouncements, etc. The corporate image is something fluid that can change overnight from positive to negative and the other way around to neutral. Large firms but also SME's use various corporate advertising techniques to enhance their image in order to improve their desirability as a supplier, employer, customer, borrower, etc. 

Incidents and troubles like the documented in this work have a great impact in the involved firms’ public perception. This is why history shows that every company involved in issues with their robots try to avoid its impact in the public sphere at almost any cost. When journalists do their job insisting and researching and the problems are communicated or leaked to media, insecurity affects the companies image. Both internally –among employees– and externally –among customers, policy makers, investors–. This has happened in almost every documented case in the industrial sector. When the Kawasaki factory worker was killed by a robot back in 1981, details of the accident were not revealed until December, at least 5 months after, by the Labour Standards Bureau of Hyogo prefecture, in western Japan. But even back in 2015, in the age of communication and preached transparency, it took Volkswagen's communication department more than ten days to openly admit that a robot killed a worker in one of their plants. To be precise, and according to the local newspaper Hessische Niedersächsische Allgemeine Zeitung (HNA), the multinational did only speak up when the journalists started inquiring \cite{mediavw}. 

The consequences for the corporate image of the firm going trough a situation like this can be a disaster in terms of Public Relations (PR). Following with the same example, immediately after the Volkswagen incident was revealed, and at a great speed, news around the world became fraught with the case of a worker being killed by a robot in a car factory. The story was picked up in the Washington Post, CNN or even in Daily Pakistan. Shortly thereafter, the American company HBO planned to shoot a documentary entitled 'Asimov's Law' on the relationship between humans and robots, which is to be published in 2019.

\section{Conclusion}
\label{sec:conclusion}
We are witnessing the dawn of robotics. But lessons should be learned from previous technological revolutions, such as the computer industry or the smart-phone revolution. We aim  to create awareness about the need of caring not only about safety when deploying a robot, but also enforcing strong security in robots. Particularly, we foresee cobots as the rising point of are the robotics technological revolution, called to change many industrial operations but also to influence greatly  the professional and consumer sectors. In this context, we find extremely relevant to promptly alert about the imperative of ensuring a security first approach now, before new devices continue being irresponsibly rushed to the market by a number of companies. Likewise, we encourage all sides involved to develop both internal and external policies aimed at the adequate management of some of the safety and security issues highlighted throughout this document.

How long will it take until manufacturers notice the damages that the insecurity of these devices may cause and act? Consequences of not facing the problem range from data theft and privacy issues, or destruction of the corporate image, to human loss and injuries, beyond their implications within any economical dimension. And yet, it seems decisive to remind that, in a world where IT and robotics are increasingly intertwined, safety and security are necessarily tightly coupled. A security-first approach is a "sine qua non" requisite for ensuring safe operations with robots. 

Robot hazards are a reality and concerns are already moving from safety to security.




\section*{Acknowledgements}

This research has been partially funded by the Basque Government, in particular, by the business development basque agency (SPRI) through the \emph{Ekintzaile} 2018 program. Special thanks to BIC Araba.


\bibliography{iclr2018_workshop}
\bibliographystyle{iclr2018_workshop}


%
%
%
%
%

\end{document}